\documentclass[journal,12pt,onecolumn, draftclsnofoot]{IEEEtran}
\usepackage[dvips]{graphicx}
\usepackage{multirow}
\usepackage{caption}
\usepackage{multicol,lipsum}
\usepackage{mathtools, cuted}
\usepackage{comment}

\usepackage{amsmath}
\usepackage{amsthm}
\usepackage[flushleft]{threeparttable}
\usepackage{array}
\usepackage{booktabs}
\newcommand{\PreserveBackslash}[1]{\let\temp=\\#1\let\\=\temp}
\newcolumntype{C}[1]{>{\PreserveBackslash\centering}p{#1}}
\newcolumntype{R}[1]{>{\PreserveBackslash\raggedleft}p{#1}}
\newcolumntype{L}[1]{>{\PreserveBackslash\raggedright}p{#1}}

\usepackage{bm}
\usepackage{graphicx}
\usepackage{caption}
\usepackage{graphicx}
\usepackage{subcaption}
\usepackage{}
\usepackage{algorithmic, algorithm}
\usepackage{cite}
\usepackage{amsmath}
\usepackage{amssymb}
\usepackage{psfrag}
\usepackage{empheq}
\usepackage{latexsym, amsmath, color, amsfonts, amssymb,graphicx}
\usepackage{wrapfig} 

\IEEEoverridecommandlockouts
\usepackage{tabularx}
\makeatletter
\def\hlinewd#1{%
\noalign{\ifnum0=`}\fi\hrule \@height #1 %
\futurelet\reserved@a\@xhline}
\makeatother
\begin{document}
\title{Distributed Deep Convolutional Compression for Massive MIMO CSI Feedback}
\author{Mahdi Boloursaz Mashhadi, Qianqian Yang, and Deniz G\"{u}nd\"{u}z\\
\IEEEauthorblockA{Dept. of Electrical and Electronic Eng., Imperial College London, UK\\
Email: \{m.boloursaz-mashhadi, q.yang14, d.gunduz\}@imperial.ac.uk}}

\maketitle
\begin{abstract}
Massive multiple-input multiple-output (MIMO) systems require downlink channel state information (CSI) at the base station (BS) to achieve spatial diversity and multiplexing gains. In a frequency division duplex (FDD) multiuser massive MIMO network, each user needs to compress and feedback its downlink CSI to the BS. The CSI overhead scales with the number of antennas, users and subcarriers, and becomes a major bottleneck for the overall spectral efficiency. In this paper, we propose a deep learning (DL)-based CSI compression scheme, called \textit{DeepCMC}, composed of convolutional layers followed by quantization and entropy coding blocks. In comparison with previous DL-based CSI reduction structures, DeepCMC proposes a novel fully-convolutional neural network (NN) architecture, with residual layers at the decoder, and incorporates quantization and entropy coding blocks into its design. DeepCMC is trained to minimize a weighted rate-distortion cost, which enables a trade-off between the CSI quality and its feedback overhead. Simulation results demonstrate that DeepCMC outperforms the state of the art CSI compression schemes in terms of the reconstruction quality of CSI for the same compression rate. We also propose a distributed version of DeepCMC for a multi-user MIMO scenario to encode and reconstruct the CSI from multiple users in a distributed manner. Distributed DeepCMC not only utilizes the inherent CSI structures of a single MIMO user for compression, but also benefits from the correlations among the channel matrices of nearby users to further improve the performance in comparison with DeepCMC. \color{black}We also propose a reduced-complexity training method for distributed DeepCMC, allowing to scale it to multiple users, and suggest a cluster-based distributed DeepCMC approach for practical implementation.\color{black} 

\makeatletter{\renewcommand*{\@makefnmark}{}\footnotetext{This work was supported by the European Research Council (ERC) through project BEACON (grant no 677854). Part of this work was presented at the IEEE 29th International Workshop on Machine Learning for Signal Processing (MLSP), Pittsburg, PA, Oct. 2019 \cite{yang2019deep2}.}\makeatother}

\end{abstract}
\section{Introduction}
Massive multiple-input multiple-output (MIMO) systems are considered as the main enabler of 5G and future wireless networks thanks to their ability to serve a large number of users simultaneously, achieving impressive levels of energy and spectral efficiency. The base station (BS) in a massive MIMO setting relies on the downlink channel state information (CSI) to fully benefit from the available degrees of freedom and achieve the promised performance gains \cite{MMIMO1, MMIMO2}. In time division duplex (TDD) mode of operation, massive MIMO systems can exploit the uplink CSI for downlink transmission, thanks to channel reciprocity. On the other hand, frequency division duplex (FDD) operation is more desirable due to the better coverage it provides; however, channel reciprocity does not hold in FDD; and hence, downlink CSI is estimated at the user and fed back to the BS.

The resulting feedback overhead becomes excessive due to the massive number of antennas and users being served, and has motivated various CSI reduction techniques based on vector quantization \cite{CodebookCSI1} and compressed sensing (CS) \cite{CSCSI1, CSCSI2}. In vector quantized CSI feedback, the overhead scales linearly with system dimensions, which becomes restrictive in many practical massive MIMO scenarios. On the other hand, CS-based approaches rely on sparsity of the CSI data in a certain transform domain, which may not represent the channel structure accurately for many practical MIMO scenarios. CS-based approaches are also iterative, which introduces additional delay.

Following the recent resurgence of machine learning, and more specifically deep learning (DL) techniques for physical layer communications \cite{Oshea2017, MLintheAir}, DL-based MIMO CSI estimation, compression and feedback techniques have recently been proposed \cite{Overview1, Overview2}. The DL-based CSI compression scheme, CSINet \cite{wen2018deep}, showed significant improvement over previous works that utilized compressive sensing and sparsifying transforms. Following CSINet, several subsequent schemes were proposed which use autoencoder architectures to reduce the MIMO CSI feedback overhead by learning low-dimensional features of the channel gain matrix from training data \cite{wen2018deep, DLCSI2, DLCSI3, DLCSI4, Access1, mashhadi2019cnnbased, FEDDEL, CSINETPlus, CRNet, lu2019bitlevel, liu2019efficient}. In \cite{DLCSI3}, the  authors improve CSINet by utilizing a recurrent neural network to utilize temporal correlations in time-varying channels. Utilizing bi-directional channel reciprocity, the authors in \cite{DLCSI4} use the uplink CSI as an additional input to further improve the results utilizing the correlation between downlink and uplink channels.

Aforementioned CSI reduction techniques focus on dimensionality reduction by direct application of the autoencoder architecture. These works are based on the assumption that reducing the dimension of the CSI matrix to be fed back to the BS would result in reduced feedback overhead. However, in general, the reduced dimension CSI matrix does not result in the most efficient representation, and it can be further compressed by efficient quantization and compression techniques. Design of efficient compression techniques and the impact of such compression on the CSI reconstruction accuracy has not been considered in \cite{wen2018deep, DLCSI2, DLCSI4}. The authors in \cite{lu2019bitlevel} use uniform quantization on the reduced CSI values. However, the distribution of the output of the encoder neural network is not uniform, and uniform quantization produces values that are not equally probable, and can be further compressed. Considering this, the authors in \cite{CSINETPlus} use non-uniform $\mu$-law quantization to get more evenly distributed quantized symbols. More recently, DL-based architectures are proposed in \cite{shlezinger2019deep} and \cite{liu2019efficient} to learn a non-uniform quantizer.

In this paper, we propose a DL-based CSI compression scheme, called DeepCMC, composed of a novel fully convolutional autoencoder structure, employing residual layers at the decoder for more accurate reconstruction, in conjunction with quantization and entropy coding blocks, which allow us to approach the fundamental limits of compression more closely. More specifically, this is the first work on MIMO CSI compression that uses an estimate of the probability distribution of the quantized autoencoder output to efficiently compress it by a context-adaptive arithmetic entropy coder at rates closely approaching its entropy. Following our initial work, arithmetic entropy coding is also adopted by \cite{liu2019efficient} for CSI compression. Here, we also propose a novel distributed DeepCMC architecture to encode the CSI from multiple users in a distributed manner, which are decoded jointly at the BS. Our goal is to exploit the correlations among the CSI matrices of nearby users to further reduce the required communication overhead. Note that a major benefit of a massive MIMO BS is its ability to simultaneously serve a large number of users in its coverage area. This means that users/devices will be located within close physical proximity of each other; and hence, exploiting common structures and correlations among their channel matrices, to better compress their CSI can significantly improve the overall spectral efficiency by reducing the resources dedicated to CSI feedback. 

\color{black}In comparison with the previous DL-based CSI compression techniques, the main contributions of the proposed DeepCMC architecture and its distributed version can be summarized as follows:

i) Existing DL-based architectures for CSI compression all include a fully connected layer, which means that they can only be utilized for a specified input size, e.g., for a given number of OFDM sub-carriers. This would mean that a different NN needs to be trained for every different resource allocation setting, and users need to store NN coefficients for all these networks, limiting the practical implementation of these solutions. Instead, the proposed DeepCMC architecture is fully convolutional, and has no densely connected layers, which makes it flexible for a wider range of MIMO scenarios. \color{black}Our simulations show that the convolutional kernels of DeepCMC, once trained, work sufficiently well for a large range of sub-carriers and antennas.

ii) Many of the existing DL-based architectures for CSI compression focus on dimensionality reduction by direct application of the autoencoder architecture and do not consider further compression of the CSI at a bit level \cite{wen2018deep, DLCSI2, DLCSI4}. DeepCMC includes quantization and entropy coding blocks within its architecture to directly convert the channel gain matrix into bits for subsequent communication. In contrast to previous works that minimize the reconstruction mean square error (MSE) of the reconstructed CSI matrix, DeepCMC minimizes a weighted rate-distortion cost that takes into account both the compression rate (in terms of bits per CSI value) and the reconstruction MSE, which significantly improves the performance and enables a rate-distortion trade-off. Although uniform and non-unifrom $\mu$-law quantization are considered in \cite{lu2019bitlevel} and \cite{CSINETPlus}, respectively, the quantization process is still blind to the specific distribution of the reduced CSI values. However, our proposed DeepCMC scheme learns the local probability distributions of the quantizer output and uses it in conjunction with context-adaptive arithmetic entropy coding to efficiently compress the quantizer output at rates closely approaching its entropy. \color{black}We provide an ablation study to evaluate the improvements due to our proposed convolutional feature encoder/decoder architecture and the use of entropy coder for compression, separately.

iii) We propose distributed DeepCMC for a multi-user massive MIMO scenario such that different users compress their CSI in a distributed manner while the BS jointly reconstructs the CSI of multiple users from the received feedback messages. This is motivated by the information theoretic results on distributed lossy compression of correlated sources \cite{elgamal_kim_2011}, and is based on the fact that the CSI of nearby users are correlated as they share common multi-path components from scatterers located far away from them. Hence, distributed DeepCMC not only utilizes the inherent structures of a single MIMO channel for compression, but also benefits from the channel correlations among nearby MIMO users to further improve the performance. \color{black}Moreover, to address practical implementation issues regarding scaling of DeepCMC to the multiple user case, we propose a reduced-complexity training scheme without sacrificing the compression efficiency much. Finally, we propose a cluster-based distributed DeepCMC approach for practical implementation.

\color{black}In parallel with our work, Guo et. al. also considered the distributed CSI compression problem in \cite{Guo}, where they jointly reconstruct the CSI from two users at the BS. Their approach is different from ours as they compress the magnitude and phase of the CSI separately and use a separate decoder module at the BS to reconstruct the CSI information shared by the two users. Instead, our proposed distributed DeepCMC architecture uses summation-based information fusion branches at different locations of the joint feature decoder to add the available side-information from all the users together. These summation-based fusion branches exploit the nature of the channel gains, which comprise of the summation of multi-path signal components.\color{black}   

This paper is organized as follows. In Section II, we present the system model. In Sections III and IV we present our proposed DeepCMC scheme for massive MIMO CSI compression and its distributed version, respectively. Section V provides the simulation results and Section VI concludes the paper.


\section{System Model}\label{sec1}
We consider a massive MIMO setting, in which a BS with $N_t$ antennas serves $K$ single-antenna users utilizing orthogonal frequency division multiplexing (OFDM) over $N_c$ subcarriers. We denote by $\mathbf{H}^k\in \mathbb{C}  ^{N_c \times N_t}$ the downlink channel matrix for user $k$, and by $\mathbf{v}^k \in \mathbb{C}^{N_t \times 1}$ the precoding vector used for downlink transmission to user $k$. The received signal at user $k$ is given by
\begin{align}
    \mathbf{y}^k= \mathbf{H}^k \mathbf{v}^k x^k + \mathbf{H}^k \sum_{i \ne k} \mathbf{v}^i x^i + \mathbf{z}^k,
\end{align}
where $x^k \in \mathbb{C}$ are the data-bearing symbols, and $\mathbf{z}^k \in \mathbb{C}^{N_c \times 1}$ is the additive noise vector, for $k \in [K]\triangleq \{1, ..., K\}$. In order to design the precoding vectors $\mathbf{v}^k$ for efficient transmission, the BS requires estimates of the downlink CSI matrices, $\mathbf{H}^k$. In an FDD system, each user estimates its downlink CSI matrix through pilot-based training, and transmits the estimated CSI back to the BS. Hence, the overhead for CSI feedback from the users grows with $K \times N_c \times N_t$, and becomes prohibitive for wideband massive MIMO systems when $K$, $N_c$ and $N_t$ are large.

To cope with this challenge, the users need to efficiently compress their channel matrices $\mathbf{H}^k$. Let $\mathbf{H}^k = [\mathbf{h}^k_1, \mathbf{h}^k_2, \hdots, \mathbf{h}^k_{N_c}]^T$, where $\mathbf{h}^k_n \in \mathbb{C}^{N_t}$ is the channel gain vector of user $k$ over subcarrier $n$, $n \in [N_c]$. Assume that the BS is equipped with a uniform linear array (ULA) with response vector $\mathbf{a}(\phi)=[1, e^{-j\frac{2\pi d}{\lambda} \sin{\phi}}, \cdots, e^{-j\frac{2\pi d}{\lambda} (N_t -1) \sin{\phi}}]^T$, where  $\phi$ is the angle of departure (AoD), and $d$ and $\lambda$ denote the distance between adjacent antennas and carrier wavelength, respectively. The channel gain vectors are a summation of multipath components as 


\begin{align}\label{Hform}
    \mathbf{h}^k_n=\sqrt{\frac{N_t}{L^k}}\sum^{L^k}_{l=1} \alpha_l^k e^{-j2\pi \tau_l^k f_s \frac{n}{N_c}} \mathbf{a}(\phi^k),
\end{align}
where $L^k$ is the number of downlink multipath components for user $k$ with $\tau_l^k$ and $\alpha_l^k$ denoting the corresponding delay and propagation gain for the components, respectively, and $f_s$ is the sampling rate. According to (\ref{Hform}), the CSI values for nearby sub-channels, antennas and users are correlated due to similar propagation paths, gains, delays and AoDs. This correlation can be exploited to compress the CSI and reduce the feedback overhead. 

Designing practically efficient codes for lossy compression is challenging even for memoryless sources with explicitly defined distribution models. Here, we take an alternative data-driven approach and propose a deep NN architecture, called DeepCMC, which learns the compression scheme when trained over large datasets of channel matrices. DeepCMC uses CNN layers and entropy coding blocks to learn the CSI compression scheme that can best leverage the underlying correlations. 

For the general case of $K$ users, we have a multi-terminal lossy source coding problem \cite{elgamal_kim_2011}, where our goal is to compress correlated CSI matrices from different users in a distributed manner and at an acceptable distortion and complexity. As opposed to the single user setting, this problem is elusive even in the ideal information theoretic setting. The general solution is known only for jointly Gaussian source distributions under squared error distortion \cite{elgamal_kim_2011, Wyner}, or for discrete memoryless sources under log-loss as the distortion measure \cite{Courtade}. Here, we propose a NN architecture, called distributed DeepCMC, and train it over a large dataset of channel matrices to achieve a distributed CSI compression scheme in a data-driven manner without explicit knowledge of the underlying distributions. Distributed DeepCMC leverages the correlations among the CSI of multiple users to further improve the rate-distortion performance in comparison with separate DeepCMC architectures for each user.

\section{DeepCMC}\label{section:2}

In this section, we present our proposed NN architecture, DeepCMC, for encoding and subsequent reconstruction of downlink CSI for a single massive MIMO user. This will be extended to the multiple-user MIMO scenario in Section IV. 
\begin{figure*}[t!]
\centering
\includegraphics[scale=0.55]{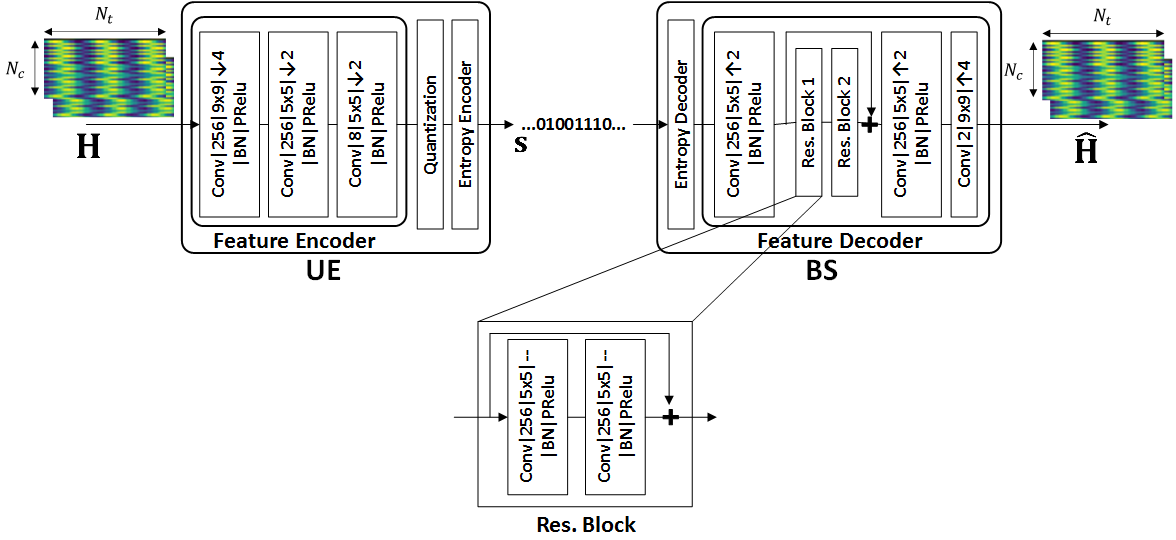}
\caption{The encoder/decoder architecture for the proposed CSI feedback compression scheme, DeepCMC.}
\label{single}
\end{figure*}
The overview of our proposed model architecture for DeepCMC is shown in Fig.~\ref{single}, where the two channel inputs represent the real and imaginary parts of the channel matrix. The user compresses its CSI into a variable length bit stream. The encoder comprises a CNN-based feature encoder, a uniform element-wise scalar quantizer, and an entropy encoder. The feature encoder extracts key features from the CSI matrix to obtain a lower dimensional representation, which is subsequently converted into a discrete-valued vector by applying scalar quantization. While previous works simply send the 32-bit scalar quantized version of the feature vector as the CSI feedback \cite{wen2018deep, DLCSI2, DLCSI4}, we have observed that the autoencoder structure does not produce uniformly distributed feature values, and hence, can be further compressed. 

To further reduce the required feedback, we employ an entropy encoder; in particular, we use the context-adaptive binary arithmetic coding (CABAC) technique  \cite{marpe2003context}, which outputs a variable-length bit stream. Upon receiving this CSI-bearing bit stream, the BS first processes it by an entropy decoder to reproduce the lower-dimensional representation of the CSI feedback which is then used by our proposed feature decoder to reconstruct the channel gain matrix. We present each component of our proposed model in more details below. 

\subsection{Feature encoder and decoder}

Fig. \ref{single} depicts the our proposed CNN architecture for the feature encoder and decoder in DeepCMC, where ``Conv$|256|9\times9|\downarrow 4|$BN$|$PReLU" represents a convolutional layer with 256 kernels, each of size $9\times9$ followed by downsampling by a factor of 4, batch normalization and parametric rectified linear unit (PReLU) activation. The feature encoder consists of three convolutional layers, the first of which uses kernels of size $9\times9$, and the other two use kernels of size $5\times5$. The ``SAME'' padding technique is used, such that the input and output of each convolutional layer have the same size (the number of channels vary). Let $\mathbf{M}=f_{\rm{f-en}}(\mathbf{H},\mathbf{\Theta}_{en})$,
where $f_{\rm{f-en}}$ denotes the feature encoder at the user, and $\mathbf{\Theta}_{en}$ denotes its parameter vector. $\mathbf{M}$ consists of $256$ feature maps of size $\frac{N_t}{16} \times \frac{N_c}{16}$. Note that this fully convolutional architecture allows us to use the same encoder network for any number of transmit antennas and subcarriers, while the feature vector dimension depends on the input size, which allows us to scale the CSI feedback volume with the channel dimension.


The feature decoder at the BS performs the corresponding inverse operations, consisting of convolutional and upsampling layers. At the BS, the output of the entropy decoder is fed into the feature decoder to reconstruct the channel gain matrix. Similarly to the feature encoder, the decoder includes three layers of convolutions (with the same kernel sizes as the encoder) and upsampling (inverse of the downsampling operation at the encoder). The decoder architecture also includes two residual blocks with shortcut connections that skip several layers with $+$ denoting element-wise addition in Fig. \ref{single}. This structure eases the training of the network by preventing vanishing gradient along the stacked non-linear layers~\cite{he2016deep}. To enable this, the input and output of a residual block must have the same size. Each residual block comprises two convolutional layers (normalized using the batch norm) and uses PReLU as the activation function. Inspired by \cite{mentzer2018conditional}, we also use an identical shortcut connecting the input and output of the residual blocks, which improves the performance as revealed by the experiments. Let $\widehat{\mathbf{H}}=f_{\rm{f-de}}(\widehat{\mathbf{M}
}, \mathbf{\Theta}_{de})$
denote the output of the joint decoder, parameterized by $\mathbf{\Theta}_{de}$, and $\widehat{\mathbf{M}}$ denote the estimate of $\mathbf{M}$ provided by the entropy decoder. $\widehat{\mathbf{H}}$ denotes the reconstructed CSI matrix at the BS. 

\color{black}
\subsection{Quantization and Entropy coding}
A major contribution of our proposed model in comparison with the existing DNN architectures for CSI compression in the literature \cite{wen2018deep, DLCSI2, DLCSI4} is the inclusion of the entropy coding block, which encodes quantized CSI data into bits at rates closely approaching its entropy. Note that, as the derivative of the quantization function is zero almost everywhere, it does not allow simple optimization with gradient descent. As a common practice in training NNs in the presence of a quantizer \cite{balle2016end, balle2018variational}, we replace the quantization and entropy coding blocks with independent identically distributed (iid) noise during training, but include them in the test phase. During training, we obtain an estimate of the quantizer output entropy in terms of the model parameters. We use this estimate as the average bit rate at the quantizer output, and add it as a term to our cost function to minimize the bit rate. We later observe in simulations that the average bit rate in the test phase closely approaches the estimated entropy, which is expected as CABAC is an efficient lossless entropy coder. 

Quantization is performed by a uniform scalar quantizer denoted by $f_{q}$. We set the quantization bin size to one, and quantize each element of $\mathbf{M}$ to the closest integer. We denote the quantized output as $\overline{\mathbf{M}}=f_{q}(\mathbf{M})$. The entropy encoder converts the quantized values in $\overline{\mathbf{M}}$ into bit streams using CABAC \cite{marpe2003context} based on the input probability distribution learned during training, denoted by $P$. More specifically, $P$ is the probability mass function of $\overline{\mathbf{M}}$  given by $P(n)=\int_{n-0.5}^{n+0.5} p_{\mathbf{M}}(x) dx, \, n \in \mathbb{Z} $ where $p_{\mathbf{M}}(x)$ denotes the probability density function of $\mathbf{M}$. Let $\mathbf{s} = f_{\rm{e-en}}(\overline{\mathbf{M}}, P)$
denote the bit stream obtained after entropy coding. Note that, as CABAC is an efficient lossless compression technique, the average bit rate at CABAC output is expected to closely follow the entropy of the quantized values $\overline{\mathbf{M}}$ given by $-\mathbb{E}[\log_2 P]$, where $\mathbb{E}[\cdot]$ denotes the expectation operator. Hence, during training we minimize this entropy term as an estimate of the average number of bits required to compress the CSI.

The estimate of $\mathbf{M}$, denoted by $\widehat{\mathbf{M}}$, is recovered at the BS by decoding the received codeword $\mathbf{s}$ using the corresponding entropy decoder as $\widehat{\mathbf{M}}= f_{\rm{e-de}}(\mathbf{s}, P)$. Note that as CABAC is a lossless compression scheme we have, $\widehat{\mathbf{M}}=\overline{\mathbf{M}}$. 
Finally, $\widehat{\mathbf{M}}$ is fed into the feature decoder to reconstruct the CSI matrix. Note that the scalar uniform quantizer followed by arithmetic entropy coding (CABAC) in our DeepCMC architecture acts as an adaptive variable bit-depth quantizer that optimally encodes the input at rates closely approaching its entropy. This alleviates the need to design more complex non-uniform quantizer blocks that optimize the quantizer thresholds according to the input distribution as proposed in \cite{shlezinger2019deep, liu2019efficient}.

\subsection{Optimization}\label{opt}
As the derivative of the quantization function is zero almost everywhere, it does not allow simple optimization with gradient descent. Similarly to \cite{balle2016end, balle2018variational}, we replace the uniform scalar quantizer with independent and identically distributed (i.i.d.) uniform noise, i.e., $\mathcal{U}[0,1]$, during training. Hence, denoting the quantization noise vector by $\Delta\mathbf{M}$ with i.i.d. elements from  $\mathcal{U}[0,1]$, we approximate the quantized feature matrix by $\widetilde{\mathbf{M}}=\mathbf{M}+\Delta\mathbf{M}$. With this simple replacement, the probability density function of $\widetilde{\mathbf{M}}$ is a continuous relaxation of the probability mass function of $\overline{\mathbf{M}}$, where $p_{\widetilde{\mathbf{M}}}(n)=P(n), \, n \in \mathbb{Z}$; and hence, we use the differential entropy of $\widetilde{\mathbf{M}}$ as an approximation of the entropy of $\overline{\mathbf{M}}$ in the cost function.  

We denote by $p_{\widetilde{\mathbf{M}}}(x,\mathbf{\Theta}_{p})$ the probability density function of $\widetilde{\mathbf{M}}$ specified by the set of parameters $\mathbf{\Theta}_{p}$, which is estimated through training similarly to \cite{balle2018variational}. Similarly to \cite{balle2018variational}, we model the cumulative distribution function of $\widetilde{\mathbf{M}}$ as a composition of $K$ functions $\{f_k\}_{k=1}^{K}$, each of which is modeled by a NN as $f_k(x)=\sigma_k(H_k x+b_k)$, where $H_k$ and $b_k$ are trainable parameters and $\sigma_k$ denotes the non-linearity. We refer the reader to [31, Section 6.1] for more details on the choice of $\sigma_k$'s. Hence, $p_{\widetilde{\mathbf{M}}}(x,\mathbf{\Theta}_{p})=f_K' \times f_{K-1}' \times ... \times f_1'$, where $\mathbf{\Theta}_{p}$ denotes the set of trainable parameters $\{H_k, b_k\}_{k=1}^K$. Having optimized these parameters during training, we obtain $P(n)=p_{\widetilde{\mathbf{M}}}(n), \, n \in \mathbb{Z}$ which is then used by CABAC during inference to encode the quantized values into bits and decode the bits back to values.\color{black}

Our loss function is given by
\begin{align}\label{Loss}
L(\mathbf{\Theta}_{en}, \mathbf{\Theta}_{de}, \mathbf{\Theta}_{p}) & = \mathbb{E}_{\mathbf{H}, \Delta \mathbf{M}}\Bigg(-\frac{1}{N_cN_t}\log p_{\widetilde{\mathbf{M}}}(f_{\rm{f-en}}(\mathbf{H}, \mathbf{\Theta}_{en})+\Delta \mathbf{M}, \mathbf{\Theta}_{p}) \nonumber \\ 
& + \lambda \text{MSE} \bigg(f_{\rm{f-de}}\big(f_{\rm{f-en}}(\mathbf{H}, \mathbf{\Theta}_{en})+\Delta \mathbf{M}, \mathbf{\Theta}_{de}\big), \mathbf{H}\bigg) \Bigg),   
\end{align}
where  
\begin{equation}\label{Loss1}
\text{MSE} \bigg(\widehat{\mathbf{H}}, \mathbf{H}\bigg)=\frac{1}{N_cN_t} \|\mathbf{H}-\mathbf{\hat{H}}\|_2^2, \nonumber
\end{equation}
and the expectation is over the training set of channel matrices and the quantization noise. During training, the entropy of the quantizer outputs, estimated by the trainable probability model, is jointly minimized with the reconstruction MSE by optimizing the parameters for both the probability model and the autoencoder. By utilizing the entropy coding block with the optimized probability model, the actual bit rate of the encoder output closely approximates this entropy. More precisely, the first part of the loss function in (\ref{Loss}) represents the entropy of the feedback data, or equivalently the size of the feedback in bits that must be transmitted, while the second part is the weighted MSE of the reconstructed channel gain matrices. Hence, training $\mathbf{\Theta}_{en}, \mathbf{\Theta}_{de}$ and $\mathbf{\Theta}_{p}$ values, which parameterize the feature encoder, the feature decoder, and the probability models, respectively, minimizes the feedback overhead and the reconstruction loss, simultaneously.

The $\lambda$ value governs the trade-off between the compression rate and the reconstruction loss. A larger $\lambda$ leads to a better reconstruction but a higher feedback overhead, and vice versa. In order to recover the trade-off between the compression rate and the reconstruction loss, we train DeepCMC with different $\lambda$ values. For a small $\lambda$ value, the network tries to reduce the feedback rate, while as $\lambda$ increases, it tries to keep the MSE under control while slightly increasing the rate. After training, each $\lambda$ value specifies a set of parameters $\mathbf{\Theta}_{en}, \mathbf{\Theta}_{de}, \mathbf{\Theta}_{p}$. By selecting the $\lambda$ value according to user's requirements in terms of CSI quality and the available feedback capacity, we can obtain the encoder and decoder parameters with the best performance under these constraints. This would require the user and the BS to have a list of encoder/decoder parameters to be used for different rate-MSE quality trade-offs, and the user to send the $\lambda$ value together with the encoded bitstream $\mathbf{s}$ to the BS, so that the BS employs the matching decoder parameters.

We emphasize here that the feature encoder and decoder networks are fully convolutional, and do not include any fully connected layers. Moreover the implemented entropy code can operate on inputs of any size. Therefore, the DeepCMC architecture can be trained on, or used for any channel matrix whose height and width are multiples of 16, since the feature encoder has a total downsampling rate of 16 (or, of any size, which can be made a multiple of 16 by padding). This is another advantage of DeepCMC with respect to existing NN-based CSI compression techniques, which are all trained for a particular input size. 

\section{Distributed DeepCMC}

\begin{figure*}[t!]
\centering
\includegraphics[scale=0.55]{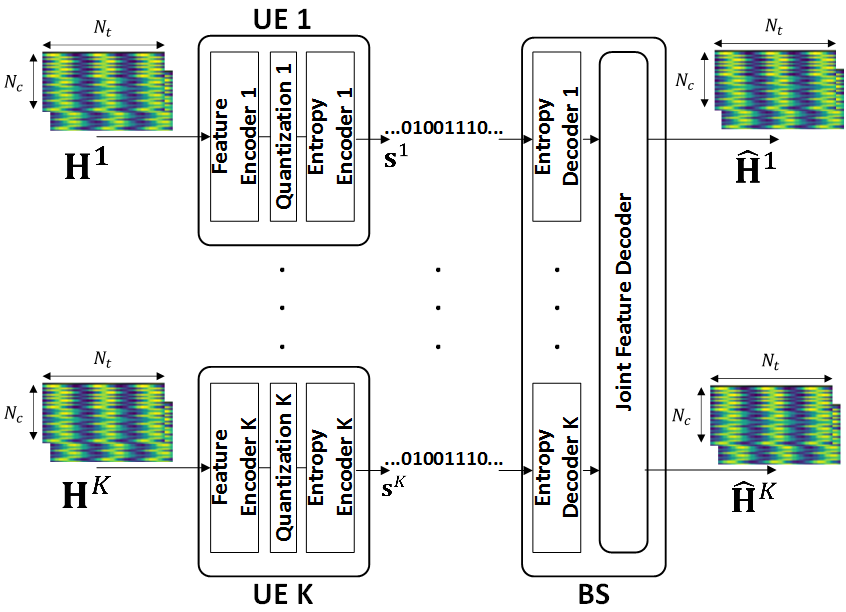}
\caption{The encoder/decoder architecture of DeepCMC for multiple-user scenario.}
\label{multi}
\end{figure*}

In a multi-user FDD massive MIMO scenario with $K$ users, each user needs to compress and feedback its downlink CSI to the BS, separately. However, if the users are located close to each other, we expect their CSI matrices to be correlated as they share some common multipath components. Even though the compression is carried out separately at the users, they can benefit from the correlation among their CSI matrices to achieve a better trade-off between the compression rate and the reconstruction MSE if the BS jointly reconstructs the CSI of multiple users from the received feedback messages. This is motivated by the information theoretic results on distributed lossy compression of correlated sources \cite{elgamal_kim_2011}. To this end, we propose a distributed DeepCMC NN architecture in which a joint feature decoder is used to simultaneously reconstruct the CSI matrix for several users at the BS. 

Fig. \ref{multi} provides the overall block diagram of the proposed distributed DeepCMC architecture. According to this figure, a $K$ user distributed DeepCMC architecture consists of $K$ separate encoder branches each consisting of a feature encoder, quantization and entropy encoder blocks to compress the downlink CSI from users to $K$ bitstreams. The feature encoder, quantization and entropy encoder block architectures are the same as described for the single user DeepCMC architecture. At the joint decoder, the bitstreams go through $K$ separate entropy decoders with the same architecture as described in the previous section. The output of the entropy decoders are input to the joint feature decoder. 
\color{black}

\begin{figure}[t!]
\centering
\includegraphics[scale=0.55]{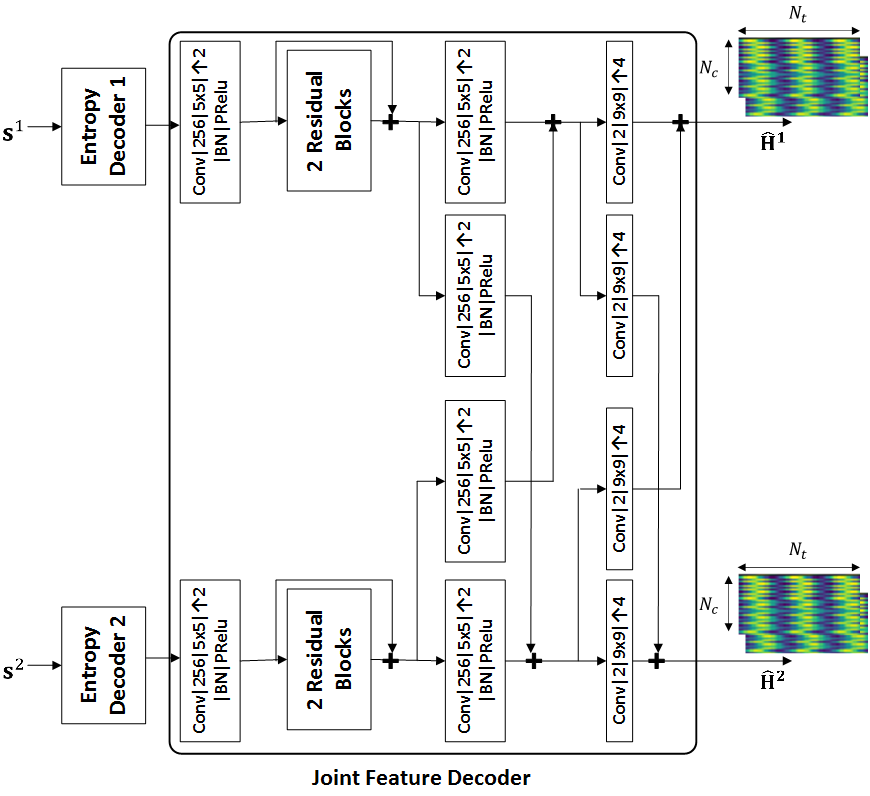}
\caption{Joint feature decoder architecture.}
\label{JFD}
\end{figure}

\subsection{Multi-user Information Fusion}
To design the joint feature decoder block, consider downlink CSI matrices of two nearby users denoted by $\mathbf{H}^{1}=[\mathbf{h}^1_{1}, \mathbf{h}_2^{1}, \hdots, \mathbf{h}_{N_c}^{1}]^T$ and $\mathbf{H}^{2}=[\mathbf{h}_1^{2}, \mathbf{h}^2_{2}, \hdots, \mathbf{h}_{N_c}^{2}]^T$. According to (\ref{Hform}), $\mathbf{h}_n^{1}$ and $\mathbf{h}_n^{2}$ can be written as the summation of multipath components. Note that, if the two users are located close to each other, the components impinging from scatterers located far away from them appear with similar angle of arrival, gain, and delay values in their CSI matrices. Hence, $\mathbf{h}_n^{1}$ and $\mathbf{h}_n^{2}$ share similar components coming from far scatters. This motivated us to use a summation-based joint feature decoder as depicted in Fig. \ref{JFD}, for $K=2$. The input from each entropy decoder is processed in separate branches for each user. The structure of these separate branches is the same as in the single user DeepCMC except for the summation-based information fusion branches between the two users. These branches combine side information on the shared CSI components, from one user with the other, by element-wise summation of the corresponding feature values (with appropriate combining kernels). The NN learns the optimal combining kernels through training. Note that in Fig. \ref{JFD} two sets of combining branches are depicted; however, more combining connections could be used between the two branches. The number of connections and their positions affect the reconstruction performance of the joint feature decoder. We tried different architectures with more/less information fusion branches in different positions between the two users but found the current architecture as depicted in Fig. \ref{JFD} to be most effective balancing the performance with complexity. Finally, note that the architecture presented in Fig. \ref{JFD} for the joint feature decoder can be easily generalized to any number of users.  
\color{black}

\color{black}
\subsection{Multi-user Training Schemes}
We propose two training schemes for distributed DeepCMC. In the first scheme, we train the whole network from scratch. However, we observed that there is a strong similarity between the kernels trained for the individual branches in the distributed scheme and the ones trained for the single user case. Hence, in our alternative training scheme, we initialize the network parameters (including those of the feature encoders, the entropy encoders/decoders, and the individual branches in the joint feature decoder) with those optimized for the single-user case, and then fine-tune all network parameters (including the parameters mentioned above and the combination kernels) for a few more training steps to get the network parameters for the distributed case. We later observe that the fine-tuning approach significantly reduces the training complexity at negligible performance loss; and hence, can be used to improve scalability of our distributed scheme for a larger number of users.     
\color{black}

\color{black}
\subsection{Multi-user Loss Function}
Our loss function for the distributed DeepCMC is given as  follows:  

\begin{align}\label{LossD}
&L(\mathbf{\Theta}_{en}^{1:K}, \mathbf{\Theta}_{de}, \mathbf{\Theta}_{p}^{1:K}) =\mathbb{E}_{\mathbf{H}^{1:K}, \Delta \mathbf{M}^{1:K}}\Bigg(-\frac{1}{N_cN_t} \sum_{k=1}^{K} \log p_{\widetilde{\mathbf{M}}}^k(f_{\rm{f-en}}^k(\mathbf{H}^k, \mathbf{\Theta}_{en}^k)+\Delta \mathbf{M}^k, \mathbf{\Theta}_{p}^k) \nonumber \\ 
& + \sum_{k=1}^{K}\lambda_k\text{MSE} \bigg(f_{\rm{f-jde}}\big(f_{\rm{f-en}}^1(\mathbf{H}^1, \mathbf{\Theta}_{en}^1)+\Delta \mathbf{M}^1, ..., f_{\rm{f-en}}^K(\mathbf{H}^K, \mathbf{\Theta}_{en}^K)+\Delta \mathbf{M}^K, \mathbf{\Theta}_{de}\big)\big[k\big], \mathbf{H}^k\bigg) \Bigg),
\end{align}
where the superscript $k$ specifies the corresponding user and $\mathbf{X}^{i:j}$ denotes the sequence $\mathbf{X}^i, \mathbf{X}^{i+1}, \hdots,$ $ \mathbf{X}^{j}$. In particular, $f_{\rm{f-en}}^k(\cdot, \mathbf{\Theta}_{en}^k)$ denotes the feature encoder at user $k$, parameterized by $\mathbf{\Theta}_{en}^k$, $\Delta\mathbf{M}^k$ is the quantization noise vector with i.i.d. elements from $U[0,1]$ that is added to the feature encoder output to replace the quantization operation during training, and $p_{\widetilde{\mathbf{M}}}^k(\cdot, \mathbf{\Theta}_{p}^k)$ denotes the probability density function parameterized by $\mathbf{\Theta}_{p}^k$. The joint feature decoder is denoted by $f_{\rm{f-jde}}\big(\cdot, \mathbf{\Theta}_{de}\big)$, parameterized by $\mathbf{\Theta}_{de}$. Note that the joint feature decoder uses all the outputs from $K$ entropy decoders, and outputs the CSI reconstruction of all the $K$ users, where $f_{\rm{f-jde}}\big(\cdot, \mathbf{\Theta}_{de}\big)[k]$ denotes the reconstruction of user $k$'s channel matrix at the BS. The expectation is taken over the training set of channel matrices and the quantization noise vectors. By minimizing this loss function, the sum entropy of the feedback data from all the $K$ users (total overhead), and the weighted MSE of the reconstructed channel gain matrices are jointly minimized. Similarly to the single user case, $\lambda_k$ governs the trade-off between the feedback rate and the reconstruction quality for user $K$. A larger $\lambda_k$ results in a better reconstruction of channel matrix for user $k$ but at an increased feedback overhead. Note that non-identical values of $\lambda_1$, $\cdots$, $\lambda_K$ allows heterogeneous CSI reconstruction qualities across users. Also note that the same loss function is utilized for both of our training schemes. 
\color{black}

\begin{table}[!t] 
\centering
\caption{Performance comparison between DeepCMC and CSINet for a single-user in the indoor scenario (User randomly placed in a 20m$\times$20m square, and $N_c=256, N_t=32$).} \label{SingleIndoor}
\resizebox{12cm}{!}{%
\begin{tabular}{c|c|c|c|c|c}
Methods & $\lambda$ & Bit rate & Estimated Entropy & $\mathrm{NMSE}$ (dB)& $\rho$\\
\hline 
\multirow{5}*{DeepCMC} &$10^4$ &0.006068 & 0.003853 & -4.12 & 0.8401\\
\cline{2-6}
&$5 \times 10^4$ & 0.01353 & 0.01152 & -7.31 & 0.9337\\
\cline{2-6}
&$10^5$ & 0.02232 & 0.02094 & -8.60 & 0.9482\\
\cline{2-6}
& $5 \times 10^5$ & 0.05353 & 0.05478 & -11.83 &  0.9732\\
\cline{2-6}
& $10^6$ & 0.07658 & 0.07488 &  -12.45 & 0.9770\\
\cline{2-6}
& $5 \times 10^6$ & 0.1526 & 0.1509 & -13.57 & 0.9808\\
\hline 
\multirow{4}*{32bit CSINet} & NA & 0.015625 & NA &  -1.31 & 0.6903\\
\cline{2-6}
& NA & 0.03125 & NA & -2.90 & 0.7806\\
\cline{2-6}
& NA & 0.0625 & NA & -5.33 & 0.8856\\
\cline{2-6}
& NA & 0.15625 & NA &  -7.04 & 0.9314\\
\hline
\multirow{4}*{16bit CSINet} & NA & 0.0078125 & NA &  -1.22 & 0.6732\\
\cline{2-6}
& NA & 0.015625 & NA & -2.77 & 0.7718\\
\cline{2-6}
& NA & 0.03125 & NA & -4.56 & 0.8391\\
\cline{2-6}
& NA & 0.07815 & NA & -6.98 & 0.9354\\
\cline{2-6}
& NA & 0.15625 & NA &  -8.67 & 0.9615\\
\hline
\end{tabular}}
\end{table}

\section{Simulations}\label{sec4}
We use the COST 2100 channel model \cite{COST2100} to generate sample channel matrices for training and testing. We consider an indoor picocellular scenario at 5.3 GHz and and outdoor rural scenario at 330 MHz band. The BS is equipped with a ULA of dipole antennas at half the wavelength spacing which is positioned at the center of a $20\mathrm{m} \times 20\mathrm{m}$ and $400\mathrm{m} \times 400\mathrm{m}$ square area for the indoor and outdoor scenarios, respectively. Note that we have presented the results for both indoor and outdoor scenarios in subsection V.A, but as the simulations revealed very similar results and conclusions for both scenarios, we have only provided the simulation results for the indoor scenario in the subsequent subsections to avoid tedious discussions of similar results. We train our models on datasets of 80000 and test on 20000 CSI realizations generated by the COST 2100 model. Each CSI realization considers a random scattering environment following the default settings in \cite{COST2100}. We use the tensorflow compression library at \cite{tfc} for DeepCMC implementation.

\begin{table}[!t] 
\centering
\caption{Performance comparison between DeepCMC and CSINet for a single-user in the outdoor scenario (User randomly placed in a 20m$\times$20m square, and $N_c=256, N_t=32$).} \label{SingleOutdoor}
\resizebox{12cm}{!}{%
\begin{tabular}{c|c|c|c|c|c}
Methods & $\lambda$ & Bit rate & Estimated Entropy & $\mathrm{NMSE}$ (dB)& $\rho$\\
\hline 
\multirow{5}*{DeepCMC} &$10^4$ &0.03937 & 0.0373 & -2.23 & 0.6853\\
\cline{2-6}
&$5 \times 10^4$ & 0.05588 & 0.0541 & -4.77 & 0.8361\\
\cline{2-6}
&$10^5$ & 0.07123 & 0.0696 &    -5.98 & 0.8736\\
\cline{2-6}
& $5 \times 10^5$ & 0.09342 & 0.0917 & -6.84 &  0.9017\\
\cline{2-6}
& $10^6$ & 0.11154 & 0.1089 &   -7.35 & 0.9284\\
\cline{2-6}
& $5 \times 10^6$ & 0.15532 & 0.15331 &  -7.96 & 0.9351\\
\hline 
\multirow{4}*{32bit CSINet} & NA & 0.0625 & NA &  -1.66 & 0.7103\\
\cline{2-6}
& NA & 0.09375 & NA & -3.21 & 0.7925\\
\cline{2-6}
& NA & 0.125 & NA & -3.87 & 0.8554\\
\cline{2-6}
& NA & 0.15625 & NA &  -4.18 & 0.8671\\
\hline
\multirow{4}*{16bit CSINet} & NA & 0.0469 & NA &  -1.57 & 0.6899\\
\cline{2-6}
& NA & 0.0625 & NA &-2.91 & 0.7727\\
\cline{2-6}
& NA & 0.0781 & NA &    -3.84 & 0.8486\\
\cline{2-6}
& NA & 0.15625 & NA &  -5.23 & 0.8953\\
\hline
\end{tabular}}
\end{table}

We first present the performance of a single-user DeepCMC architecture in different scenarios in Subsection V.A, and then provide performance results for distributed DeepCMC in Subsection V.B. We use the normalized MSE (NMSE) and cosine correlation ($\rho$) as the performance measures. These measures are defined as follows:
\begin{equation}
\mathrm{NMSE} \triangleq \mathbb{E}\left\{ \frac{ \|\mathbf{H}-\mathbf{\hat{H}}\|_2^2}{ \|\mathbf{H}\|_2^2} \right\},   
\end{equation}
and
\begin{equation}
 \rho \triangleq \mathbb{E} \left\{\frac{1}{N_c} \sum^{N_c}_{n=1} \frac{|\mathbf{\hat{h}}_n^H \mathbf{h}_n|}{\|\mathbf{\hat{h}}_n\| \|\mathbf{h}_n\| }\right\}.   
\end{equation}

\begin{figure}
    \centering
    \begin{subfigure}[]{.5\textwidth}
        \centering
        \includegraphics[width=0.95\linewidth]{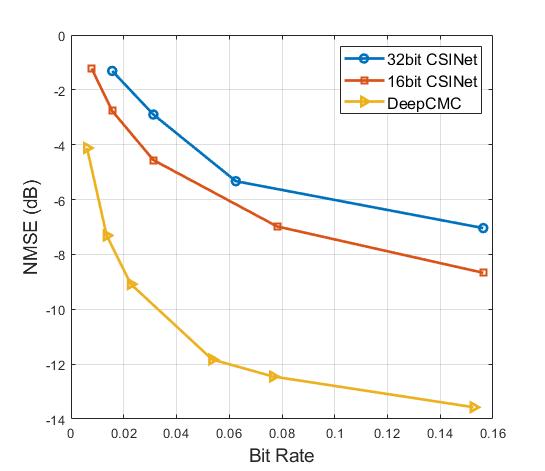}
        \subcaption{Indoor} 
        \label{csinetindoor}
    \end{subfigure}\begin{subfigure}[]{.5\textwidth}
        \centering
        \includegraphics[width=0.95\linewidth]{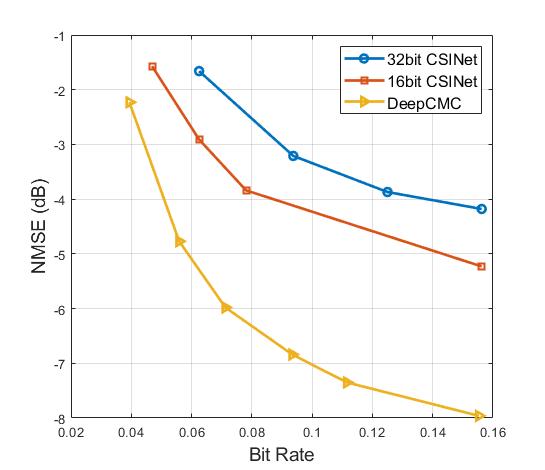}
        \subcaption{Outdoor} 
        \label{csinetoutdoor}
    \end{subfigure}

    \caption{Bit rate-NMSE trade-off of DeepCMC vs. CSINet, $N_c=256$, $N_t=32$.}
\end{figure}

\subsection{DeepCMC for a Single User}\label{SingleDeepCMC}

\color{black}
\subsubsection{Bit rate-NMSE trade-off}
We first compare the performance of our DeepCMC scheme with CSINet for the single-user scenario. We assume the user is placed uniformly at random within a $20\mathrm{m} \times 20\mathrm{m}$ and $400\mathrm{m} \times 400\mathrm{m}$ square area for the indoor and outdoor scenarios, respectively. In both scenarios, the BS is positioned at the center of the square area considered and we have $N_c=256$ and $N_t=32$. Tables \ref{SingleIndoor} and \ref{SingleOutdoor} provide the corresponding results for the indoor and outdoor scenarios, respectively. In these tables, we train our DeepCMC architecture with different $\lambda$ values, which governs the trade-off between the compression rate and the reconstruction quality. We evaluate both the average entropy of the quantized outputs of the feature encoder and the average number of actual bits to transmit back to the BS. The actual number of bits includes the length of the bit stream generated by the entropy encoder plus $16$ additional bits to transmit the value of lambda to the BS. Hence, the actual bit rate will reduce if the BS and the user agree on a fixed $\lambda$ value throughout their operation. Both the average entropy and the number of bits are normalized by $N_cN_t$, the CSI matrix dimension, to represent the average bit rate per CSI value. According to the results in Tables \ref{SingleIndoor} and \ref{SingleOutdoor}, the actual bit rate closely approximates the estimated entropy of the quantized feature encoder outputs. On the other hand, CSINet provides a feature vector of a fixed length $m$. We have considered both 32-bit and 16-bit floating point quantization for this vector. The resulting bit rate for CSINet is then given by $\frac{m\times32}{N_cN_t}$ and $\frac{m\times16}{N_cN_t}$ for the 32-bit and 16-bit CSINet, respectively.



\begin{table}[!t] \centering
\caption{Performance of DeepCMC and CSINet for a single-user in the indoor scenario (User is placed at a fixed location (5m,5m), and $N_c=256, N_t=32$).} \label{Single11}
\resizebox{12cm}{!}{\begin{tabular}{c|c|c|c|c|c}
Methods & $\lambda$ & Bit rate & Estimated Entropy & NMSE (dB)& $\rho$\\
\hline 
\multirow{5}*{DeepCMC} &100 & 0.01277 & 0.01069 & -15.02 &  0.9903\\
\cline{2-6}
&1000 & 0.03428 & 0.02868 & -23.39 & 0.998\\
\cline{2-6}
&5000 & 0.05743 & 0.05377 & -28.79 & 0.9995\\
\cline{2-6}
& 10000 & 0.06864 & 0.07709 & -31.65 &  0.9998\\
\cline{2-6}
& 50000 & 0.1163 & 0.1079 &  -37.23 & 0.9999\\
\cline{2-6}
& 100000 & 0.1459 & 0.1411 & -39.33 & 1\\
\hline 
\multirow{4}*{16-bit CSINet} & NA & 0.0078125 & NA &  -9.35 & 0.9687\\
\cline{2-6}
& NA & 0.015625 & NA & -10.07 & 0.9721\\
\cline{2-6}
& NA & 0.03125 & NA & -11.19 & 0.9766\\
\cline{2-6}
& NA & 0.07815 & NA & -12.44 & 0.9809\\
\hline
\end{tabular}}
\end{table}

The bit rate-NMSE trade-offs achieved by DeepCMC and CSINet are plotted in Figures \ref{csinetindoor} and \ref{csinetoutdoor} for the indoor and outdoor scenarios, respectively. As it can be observed from Table \ref{SingleIndoor} and Fig. \ref{csinetindoor} for the indoor scenario, DeepCMC provides significant improvement in the quality of the reconstructed CSI at the BS with respect to CSINet at both bit rate values. Note that the 16-bit CSINet performs slightly better than the 32-bit version. This is because 16-bit quantization decreases the required bit rate by a factor of two while slightly degrading the NMSE. However, we observed in simulations that further reducing the quantization precision to 8-bits or below degrades the rate-NMSE trade-off. For the indoor scenario, the NMSE values achieved by DeepCMC are around $3$ to $5$ dB lower than those achieved by the 16-bit CSINet for the range of compression rates considered in Fig. \ref{csinetindoor}. For example, for a target value of NMSE=-7dB, DeepCMC provides more than 4 times reduction in the number of bits that must be fed back from the user to the BS. Similar improvements are observed for the outdoor scenario as well. Finally, we observe from the rate-distortion curves that the NMSE values for DeepCMC drop quite rapidly with bit rate, while CSINet shows a smoother reduction slope. This implies that DeepCMC better exploits the limited number of bits to capture the most essential information in the CSI data.

These improvements are due not only to our improved feature extraction architecture, but also to the incorporation of the quantization and entropy coding blocks in the DeepCMC architecture, which enable efficient compression of the quantizer output at rates very close to its entropy. The entropy coder can efficiently convert the quantizer output to bits by utilizing its probability distribution estimated during training. Our experiments also reveal that adding the shortcut connections across two residual blocks at the decoder and choosing PReLU (in comparison with ReLU and Leaky ReLU) as the activation function improves the performance of DeepCMC.
\color{black}


\subsubsection{Stationary users}

\begin{table}[!t] 
\centering 
\caption{Performance of DeepCMC (trained with $\lambda=10^5$) in the indoor scenario for users located at different distances to the BS, $N_c=256, N_t=32$.}\label{distancetable}
\resizebox{10cm}{!}{%
\begin{tabular}{c|c|c|c|c}
Distance & Bit rate & Estimated Entropy & $\mathrm{NMSE}$ (dB) & $\rho$\\
\hline 
 $2.5\mathrm{m}$ & 0.02938 & 0.02772 & -13.33 & 0.9835 \\
\cline{1-5}
 $5\mathrm{m}$ & 0.0213 & 0.01964 & -11.01 & 0.9734\\
\cline{1-5}
 $7.5\mathrm{m}$ & 0.0192 & 0.01753 & -8.94 &  0.9586\\
\hline
 $10\mathrm{m}$ &
 0.01944 & 0.01777 & -3.56 & 0.8498\\
\hline
 random & 0.02266 & 0.02105 & -9.08 & 0.9555\\
\hline
\end{tabular}}
\end{table}

\begin{table}[!t] 
\centering 
\caption{Performance of DeepCMC (trained with $\lambda=10^5$ and $N_c=256$) for different number of subcarriers in the indoor scenario with $N_t=32$.}\label{Sizetable2}
\resizebox{9cm}{!}{%
\begin{tabular}{c|c|c|c|c}
$N_c$ & Bit rate & Entropy & $\mathrm{NMSE}$ (dB) & $\rho$\\
\hline 
 128 & 0.02493 & 0.02301 & -8.14 & 0.9469\\
\cline{1-5}
 160 & 0.02388 & 0.02217 &  -8.29 &  0.9474\\
\cline{1-5}
 192 & 0.02318 & 0.02163 & -8.42 &  0.9478\\
\hline
 224 &
 0.02269 & 0.02124 & -8.51 & 0.9480\\
\cline{1-5}
 256 & 0.02232 & 0.02094 & -8.60 & 0.9482\\
\hline
 512 & 0.021 & 0.0199 & -9.01 & 0.9490\\
\hline
 1024 & 0.02035 & 0.01938 & -9.28 & 0.9496\\
\hline
\end{tabular}}
\end{table}

For a user fixed at a certain position from the BS, we can use COST2100 to generate a dataset for that specific position and train our DeepCMC network with it. This could be the case where a wireless user is stationary (e.g., desktop PC, smart home appliances, etc. in the indoor scattering scenario) and will significantly improve the performance as there is less information in the CSI matrix of a stationary user to compress. \color{black}Note that although the user is stationary in this scenario, the scattering environment is randomly generated for each realization in the dataset, and hence, the NN still experiences random realizations of the CSI during training and testing. These random realizations differ in the number of multi-path components, their corresponding gains, delays, AoA/AoDs, etc.\color{black}

To study the performance in this scenario, we train and test both DeepCMC and 16-bit CSINet for a user fixed at $(5\mathrm{m}, 5\mathrm{m})$. Table \ref{Single11} provides the corresponding results for $N_c=256$ and $N_t=32$ in the indoor scenario. The performance gap between DeepCMC and CSINet is even larger for a fixed user. DeepCMC achieves almost perfect reconstruction with an $\mathrm{NMSE}$ of $-40$~dB and $\rho$ approximately equal to $1$ at a bit rate lower than $0.15$ bits per channel dimension for a fixed user.

\begin{figure}[!t]
\centering
\includegraphics[width=0.65\linewidth]{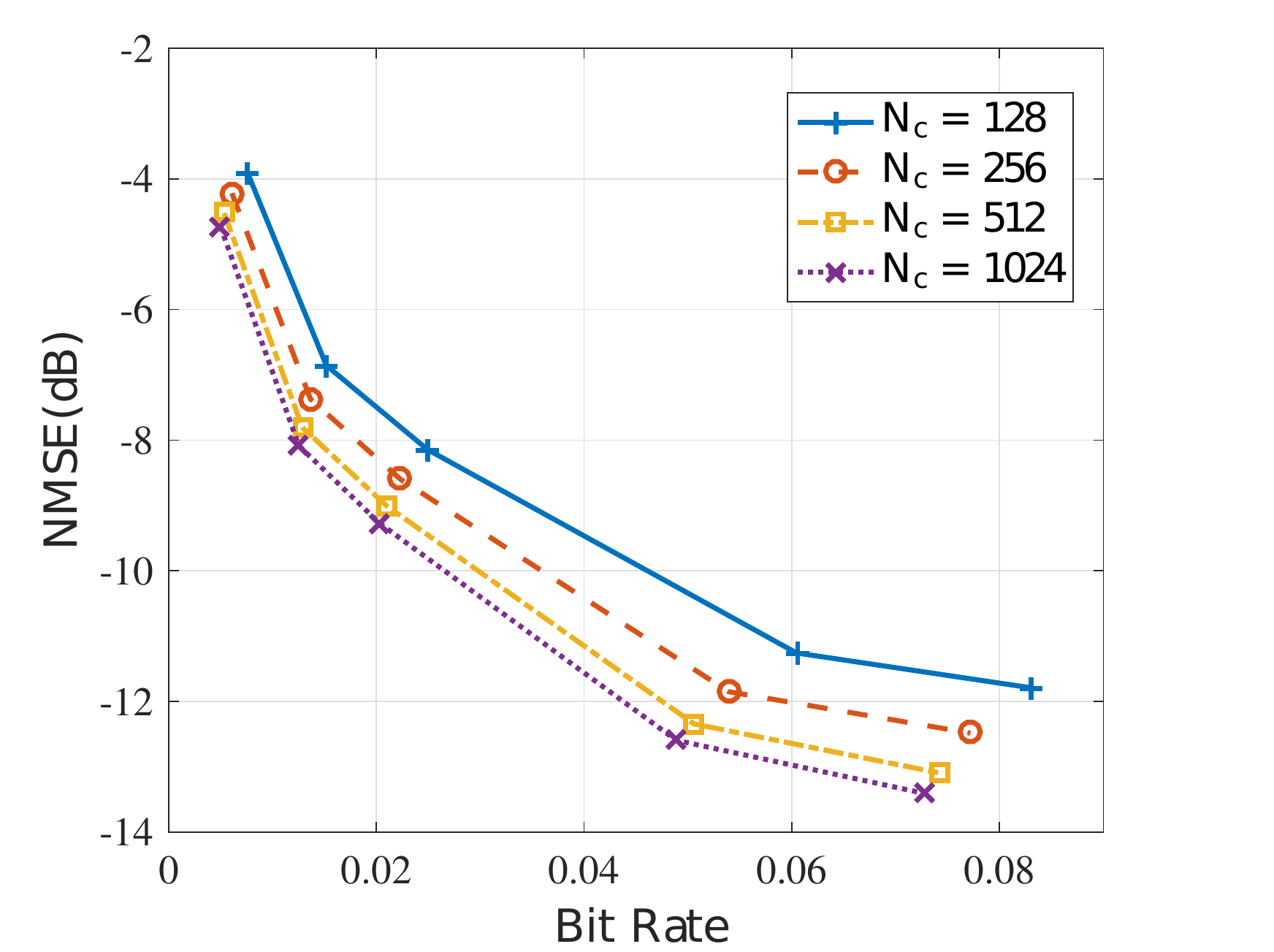}
\caption{Bit rate-NMSE trade-off for different number of subcarriers during the test phase for a DeepCMC network trained with $N_c=256$, $N_t=32$ in the indoor scenario.}\label{sizefigure}
\end{figure}

\subsubsection{User position uncertainty}
For the general scenario where the users may move, we train DeepCMC with dataset entries generated for users randomly placed in a training area. We have so far considered a $20\mathrm{m} \times 20\mathrm{m}$ square training area with the BS positioned at the center at $(10\mathrm{m}, 10\mathrm{m})$. We here study the performance of our DeepCMC network trained for the $20\mathrm{m} \times 20\mathrm{m}$ square area for users placed on circles at different distances, in particular, $2.5\mathrm{m}$, $5\mathrm{m}$, $7.5\mathrm{m}$, $10\mathrm{m}$ around the BS. We summarized the performance of DeepCMC, trained with $\lambda=10^5$, with regards to the distance between the user and the BS in Table \ref{distancetable}. The last row shows the performance when the user is randomly located within the square. Although the reconstruction performance degrades as the user moves further away from the BS, it still remains acceptable ($\mathrm{NMSE<-3 dBs}$) as long as the user stays within the training area. The NMSE for DeepCMC is smaller when the user is closer to the BS at a slightly larger bit rate. 


\begin{figure}[!t]
\centering
\includegraphics[width=0.60\linewidth]{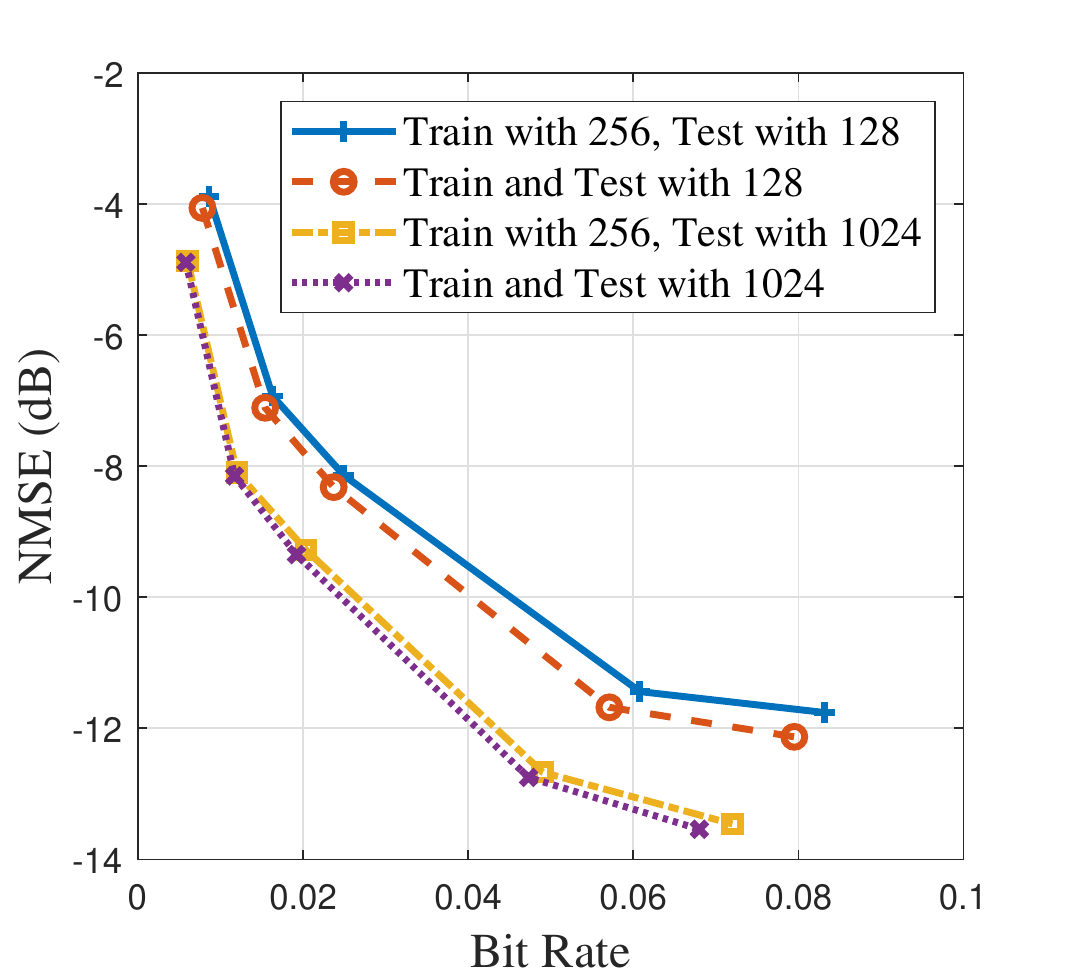}
\caption{DeepCMC performance when the number of subcarriers $N_c$, is different during the training and test phases in the indoor scenaio with $N_t=32$.}\label{sizefigure2}
\end{figure}

\subsubsection{Performance in wideband MIMO systems}

In practical MIMO scenarios, the bandwidth and consequently number of subcarriers $N_c$ may change from system to system or over time due to time-varying resource allocation. Hence, it is desirable for any CSI feedback scheme to maintain an acceptable performance as the number of subcarriers changes, so that the users will not need to store different NN parameters trained for different bandwidths. Unlike the previous works, which include dense layers in their NN architectures, DeepCMC, being fully convolutional, is applicable to scenarios with different $N_c$ values.

We design experiments to evaluate the performance of DeepCMC when trained on $N_c=256$ but tested on $N_c=128, 160, 192, 224, 256, 512, 1024$. \color{black} Note that in this experiment the carrier spacing is kept fixed at $\Delta f = 2\mathrm{MHz}$, and hence, different $N_c$ values represent systems working over different bandwidths.\color{black} We summarize the performance of DeepCMC, trained with $\lambda=10^5$, in Table \ref{Sizetable2}. We also present the bit rate-NMSE trade-off in Fig. \ref{sizefigure}, which is obtained by testing the DeepCMC (trained with different $\lambda$ values) with different values of $N_c$. According to Table \ref{Sizetable2} and Fig. \ref{sizefigure}, the DeepCMC convolution kernels once trained for $N_c=256$, work sufficiently well both on smaller and larger values of $N_c$ in a wide range of three octaves $\left( \frac{1024}{128}=8 \right)$. This is very desirable as it makes our proposed DeepCMC architecture applicable to wideband massive MIMO systems. Also according to Fig. \ref{sizefigure}, CSI matrices for wideband MIMO scenarios seem to be more compressible as larger $N_c$ values result in lower bit rate and better NMSE. 

\begin{table}[!t] 
\centering 
\caption{Performance of DeepCMC (trained with $\lambda=10^5$ and $N_t=32$) for different number of antennas in the indoor scenario with $N_c=256$.}\label{Sizetable1}
\resizebox{9cm}{!}{%
\begin{tabular}{c|c|c|c|c}
$N_t$ & Bit rate & Entropy & $\mathrm{NMSE}$ (dB) & $\rho$\\
\hline 
 16 & 0.02367 & 0.02203 & -8.32 & 0.9477\\
\cline{1-5}
 32 & 0.02232 & 0.02094 &  -8.60 &  0.9482\\
\cline{1-5}
 64 & 0.02161 & 0.01927 & -8.85 &  0.9503\\
\hline
 128 & 0.02056 & 0.01901 & -8.93 & 0.9511\\
\hline
\end{tabular}}
\end{table}

Note that, although a DeepCMC network trained on a dataset with $N_c=256$ provides very good rate-distortion curves for $N_c=128$ and $1024$ according to Fig. \ref{sizefigure}, we are interested to compare its performance with networks trained specifically on $N_c=256$ and $N_c=1024$. The corresponding comparison results are provided in Fig. \ref{sizefigure2}. According to this figure, although networks trained and tested on the same $N_c$ values provide better performance, the performance gap is small if $N_c$ is different for train and test. This shows that utilizing DeepCMC, the UE can use the kernels optimized for a specific $N_c$ value to compress the CSI for a wider range of bandwidths with negligible performance loss.

\color{black}
\subsubsection{Variation in the number of antennas}
Next, we study the flexibility of DeepCMC when the number of antennas at the test phase is different from the one in the training phase. To this end, we generate datasets for $N_t= 16, 64, 128$ (in the same ULA setting and keeping the same spacing between antennas), and test the kernels trained for $N_t=32$ (at $\lambda=10^5$) on these datasets. Table \ref{Sizetable1} reports these simulation results. According to Table VI, the kernels trained for $N_t=32$ perform sufficiently well for the range of $N_t$ values considered, which shows the flexibility of the proposed DeepCMC architecture to variations in the number of antennas during the test phase. This is thanks to the fact that, unlike CSINet, the DeepCMC architecture is fully convolutional and avoids dense layers. Larger CSI matrices (larger $N_t$ values) are also slightly better compressible. 
\color{black}

\color{black}
\subsubsection{Ablation study}

\begin{figure}[!t]
\centering
\includegraphics[width=0.6\linewidth]{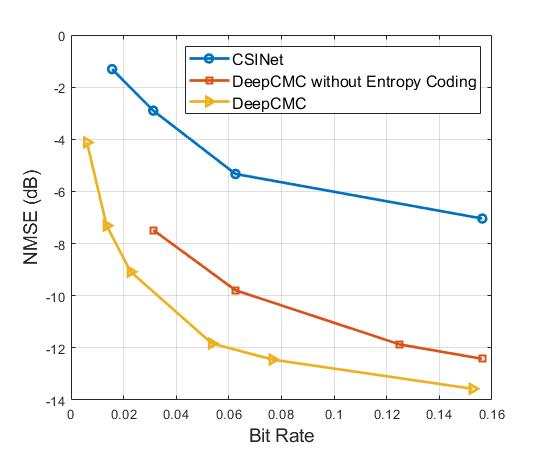}
\caption{Ablation results for $N_c=256$ and $N_t=32$ in the indoor scenario.}\label{ablation}
\end{figure}

Finally, we would like to analyze the performance gain in the DeepCMC architecture resulting from our proposed CNN architecture and the entropy coder, separately. In Fig. \ref{ablation}, we report the NMSE values resulting from our feature encoder and decoder architecture without the entropy coding block and assuming 32-bit quantized (float 32 data type) feature values, together with 32-bit quantized CSINet results. To obtain different bit-rate values for DeepCMC without the entropy coder, we change the number of convolution kernels in the last layer of the feature encoder in the range ${2, 4, 8, 10}$, and the downsample factors in the feature encoder architecture to 4, 4, 2 (accordingly, the corresponding upsample factors in the feature decoder are changed to 2,4, and 4). We see that a significant part of the improvement (in comparison with CSINet) is due to our proposed CNN-based feature encoder/decoder architecture, while the entropy coding block further reduces the NMSE. Note that a similar result holds if we compare the 16-bit quantized DeepCMC feature encoder/decoder with 16-bit CSINet.  

\color{black}

\color{black}
\subsection{Distributed DeepCMC for Two Nearby Users}
In this subsection, we present the results for the distributed implementation of DeepCMC for several users. 

\subsubsection{Effects of the training scheme}
\begin{figure}[t!]
\centering
\includegraphics[scale=.55]{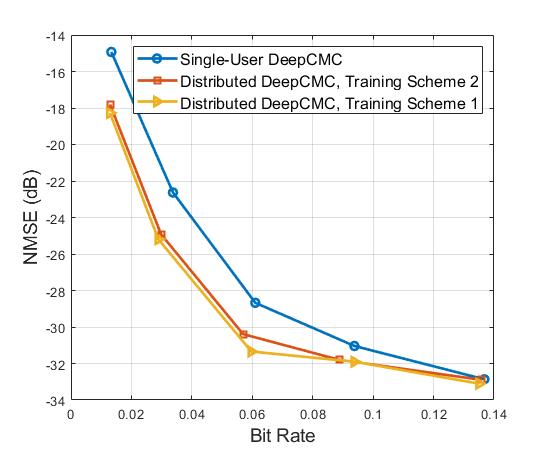}
\caption{Average NMSE vs. average bit rate achieved by single-user DeepCMC and distributed DeepCMC for the proposed training schemes (indoor scenario with two users located 60cm apart and $N_t=32$, $N_c=256$).}\label{TrainingSchemes}
\end{figure}

We consider two users placed 60cm apart in an indoor scenario with $N_c=256$ and $N_t=32$. The users are placed at locations $(5\mathrm{m}, 5\mathrm{m})$ and $(4.4\mathrm{m}, 5\mathrm{m})$, while the BS is fixed at the center $(10\mathrm{m}, 10\mathrm{m})$. As a baseline, we also consider the NMSE obtained by encoding and decoding the CSI of the two users using two independent DeepCMC networks trained seprately for each of the users. This approach does not benefit from the common structure and correlations shared by the users. Instead, in the proposed distributed DeepCMC scheme, the users encode their CSIs separately, but these CSIs are decoded jointly at the BS. For easier comparison, we plot in Fig. \ref{TrainingSchemes} the average rate and NMSEs of the two users. In this figure, Scheme 1 represents training the whole network from scratch (for 300000 steps), while Scheme 2 corresponds to initializing the network parameters to those trained for single-user DeepCMC, and fine-tuning only for 30000 more steps.

According to this figure, there is a negligible performance loss by the proposed low-complexity training Scheme 2. We highlight that to generate the results in this figure, Scheme 1 has been trained for 300000 steps while Scheme 2 is trained for only 30000 steps. Hence, our proposed fine-tuning approach significantly reduces the training complexity and time without sacrificing the performance much. This allows us to scale distributed DeepCMC to a large number of users. In the rest of this section, we only present results for distributed DeepCMC trained with the proposed low-complexity fine-tuning approach. 

\subsubsection{Effects of inter-user distance}
Next, we study the impact of the distance between the users on the performance of distributed DeepCMC. To represent typical inter-device distances in the indoor scenario in a 20m$\times$20m room, we place the users around the BS at 30cm, 60cm, and 90cm apart from each other, respectively. More specifically, we place one of the users at $(5\mathrm{m}, 5\mathrm{m})$ while the other one is placed at $(4.7\mathrm{m}, 5\mathrm{m})$, $(4.4\mathrm{m}, 5\mathrm{m})$ and at $(4.1\mathrm{m}, 5\mathrm{m})$, respectively. We use the COST2100 model to simultaneously generate CSI datasets for the two users with $N_t=32$ and $N_c=256$. We train with 80000 CSI realizations, and test over 20000 independent realizations. Other simulation parameters are the same as in Subsection \ref{SingleDeepCMC}. 

\begin{figure}[t!]
\centering
\includegraphics[scale=.55]{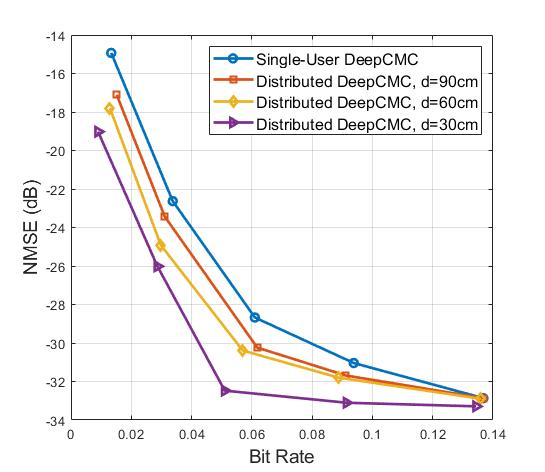}
\caption{Average NMSE vs. average bit rate for single-user and distributed DeepCMC in the indoor scenario, where the distance
between two users is 30, 60 and 90 cm ($N_t=32$, $N_c=256$).}\label{Dist}
\end{figure}

Fig. \ref{Dist} compares the average bit rate-NMSE curves achieved by distributed DeepCMC in these three cases as well as the average bit rate-NMSE curve achieved by single-user DeepCMC. We observe that distributed DeepCMC always outperforms the single-user DeepCMC, showing that the users benefit from the information transmitted by each other despite distributed encoding. The performance improvement by distributed DeepCMC becomes more significant in the low bit rate region. As expected, the improvement also increases as the users get closer to each other. This is expected as the CSI matrices for closer users have more common multipath components. For $\mathrm{d} > 100 \mathrm{cm}$, we observed no meaningful improvement by distributed DeepCMC with respect to the single-user performance. According to Fig. \ref{Dist}, for a target reconstruction NMSE of $\sim -31 \mathrm{dB}$, the average rate required for each user by DeepCMC is 0.0938 bits per channel dimension, which can be reduced to 0.0461, 0.0708, and 0.0772 by distributed DeepCMC for users located 30, 60 and 90cm apart, respectively. These correspond to 50.85\%, 24.52\%, and 17.70\% reduction in the bit rate required for CSI feedback per user, respectively. We note that these results are obtained by the simplified training scheme 2; and hence, they can be improved slightly at the expense of an increase in training complexity using scheme 1.  




\subsection{Distributed DeepCMC for More Users}
\begin{figure}[t!]
\centering
\includegraphics[scale=.55]{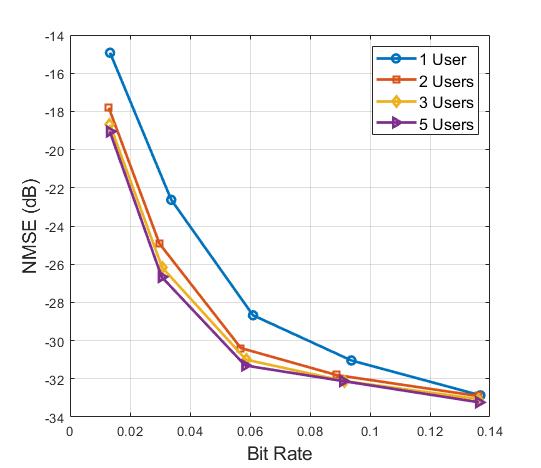}
\caption{Average NMSE vs. average bit rate for distributed DeepCMC with different number of users located equidistantly on a circle with radius $R=30 \mathrm{cm}$ (indoor scenario with $N_t=32$ and $N_c=256$).}\label{NUser}
\end{figure}

In this subsection we study the performance of distributed DeepCMC with more than two users. In particular, we consider 1, 2, 3, and 5 users located equidistantly on a circle of radius $R=$30cm centered at  $(5\mathrm{m}, 5\mathrm{m})$ in the indoor scenario ($N_t=32, N_c=256$). We expect the rate-NMSE curve to improve as we jointly decode the feedback from more users, due not only to the increased side information, but also to the reduced distance between the users. The resulting rate-NMSE curves are provided in Fig. \ref{NUser}, which reveal that the improvement becomes less significant beyond three users. 

Considering the results in Subsections V.B.2 and V.B.3, we propose a cluster-based distributed DeepCMC approach for efficient CSI feedback in practical FDD MIMO-OFDM systems. Note that the rate-NMMSE improvement by distributedDeepCMC depends on the number of users decoded jointly, their relative distances and the general system and environment-specific characteristics (e.g., carrier frequency, room geometry, BS location, number of subcarriers/antennas, etc.). For the indoor scenario considered, we have observed in Fig. \ref{Dist} that the improvement by distributed DeepCMC is more significant for users placed at distance $d<100$cm. On the other hand according to Fig. \ref{NUser}, the amount of improvement by jointly decoding additional users above 3 becomes less significant. Hence, we propose clustering the users into groups of two or three based on their location data (e.g., GPS data), such that the users in each cluster are within 1m vicinity of each other (if such clusters exist). The BS jointly reconstructs the CSI from users in the same cluster using distributed DeepCMC. With clustering, the overall complexity becomes affordable while benefiting from the most significant amount of improvement by joint decoding. The BS can default to using single user DeepCMC for all users separately. During operation of the MIMO system, whenever small clusters form (due to movements of the users or new users joining the network), the BS can switch to distributed DeepCMC to improve the overall rate-NMSE performance. 
\color{black}

\section{Conclusion}\label{sec5}
In this paper, we proposed a convolutional DL architecture, called DeepCMC, for efficient compression of CSI matrices to reduce the significant CSI feedback overhead in massive MIMO systems. DeepCMC is composed of fully convolutional layers followed by quantization and entropy coding blocks, and outperforms state of the art DL-based CSI compression techniques, providing drastic improvements in CSI reconstruction quality at even extremely low feedback rates. We also proposed a distributed version of DeepCMC for a multi-user MIMO scenario such that different users compress their CSI matrices in a distributed manner, which are reconstructed jointly at the BS. Distributed DeepCMC not only utilizes the inherent CSI structures of a single MIMO user for compression, but also benefits the channel correlations among nearby MIMO users to further improve the performance in comparison with DeepCMC. \color{black}We showed that distributed deepCMC can provide further reduction in the feedback overhead, particularly for nearby users, and proposed a low-complexity training method for distributed DeepCMC that significantly reduces the training complexity and time with only minimal performance loss.\color{black}

\bibliographystyle{IEEEtran}
\bibliography{main}
\end{document}